\documentclass[pre,twocolumn]{revtex4}
\usepackage[T1]{fontenc}
\usepackage[latin1]{inputenc}
\usepackage{graphicx}
\usepackage{amssymb}

\makeatletter

\usepackage{subfigure}
\usepackage{pslatex}
\usepackage[english]{babel}
\makeatother
\begin{document}

\title{Analytical approach to directed sandpile models on the Apollonian network}

\author{Andr\'e P. Vieira$^{1}$, Jos\'e S. Andrade Jr.$^{2,3}$, Hans J. Herrmann$^{2,3}$, and Roberto F. S. Andrade$^{4}$}
\affiliation{$^{1}$Departamento de Engenharia Metal\'{u}rgica e de
Materiais, Universidade Federal do Cear\'{a}, Campus do Pici, 60455-760,
Fortaleza, Brazil\\$^{2}$Departamento de F\'isica, Universidade Federal do
Cear\'{a}, Campus do Pici, 60455-760, Fortaleza, Brazil\\
$^{3}$Computational Physics, IfB, ETH-H\"{o}nggerberg, Schafmattstr. 6,
8093, Z\"{u}rich, Switzerland\\ $^{4}$Instituto de F\'isica, Universidade
Federal da Bahia, 40210-340, Salvador, Brazil}

\begin{abstract}
We investigate a set of directed sandpile models on the Apollonian
network, which are inspired on the work by Dhar and Ramaswamy (PRL
\textbf{63}, 1659 (1989)) for Euclidian lattices. They are characterized
by a single parameter $q$, that restricts the number of neighbors
receiving grains from a toppling node. Due to the geometry of the network,
two and three point correlation functions are amenable to exact treatment,
leading to analytical results for the avalanche distributions in the limit
of an infinite system, for $q=1,2$. The exact recurrence expressions for
the correlation functions are numerically iterated to obtain results for
finite size systems, when larger values of $q$ are considered. Finally, a
detailed description of the local flux properties is provided by a
multifractal scaling analysis.
\end{abstract}

\date{\today}
\maketitle

\section{Introduction}

The interaction networks of many real systems with large number of basic
units are often found to display power-law distribution of node degrees
and small-world property \cite{Amaral2000,Albert2002}. Examples stem from
most different areas, as electric power distribution, food webs in
ecology, information flow in the internet, interaction among financial
institutions, and so on \cite{pasvaz01,Boccaletti2006}. In recent years,
complex networks have also attracted attention as alternative topological
structures to ordered euclidian lattices, on which many physical models
can be defined. These structures offer a suitable scenario to mimic the
effect of geometry in real systems, and have already been used in the
investigation of the properties of magnetic \cite{Stauffer2002,
Dorogovtsev2004, Dorogovtsev2005} and electron \cite{Kramer2005} systems.

Understanding the stability of complex networks becomes of relevance for
the management of natural and human built systems, as it can provide
guidelines to avoid an irreversible collapse and to enhance the robustness
of their structure. Another issue that deserves attention is the
occurrence of events that may cause permanent or temporary damages on the
network, which can be interpreted as avalanches within the proposed
self-organized criticality (SOC) scenario \cite{Bak1987}. It is well known
that a typical signature of SOC systems is the possibility of occurrence
of a very large avalanche that can extend itself over the whole network,
causing its breakdown. Specific sandpile models defined on complex
networks have been recently investigated \cite{Goh2003}, as well as models
where the network is not fixed, but the set of connections evolves slowly
with time \cite{Bianconi2004}. In the first case, avalanches refer to the
motion of mass units from one node to its neighbors, while in the last
approach, avalanches refer to bursts of rewiring connections among the
network nodes. It is noteworthy the recent attempts to use SOC concepts
with respect to brain activity, both in Euclidian and scale-free
networks \cite{Arcangelis2006, Shin2006, Pellegrini2006}.

It is well known that direct models, like the one proposed by Dhar and
Ramaswamy \cite{Dhar1989}, constitute one of the few classes of SOC models
that can be exactly solved on Euclidian lattices. This is essentially
related to their Abelian property, according to which the effect of two
successive grain additions on the lattice does not depend on the order. In
the context of complex networks, the Apollonian packing problem
\cite{Herrmann1990} inspired the introduction of the so-called Apollonian
network \cite{Andrade2005,Doye2005}. Besides  displaying both scale free
and small-world features, the hierarchical geometry of this network
enables the derivation of tractable analytical expressions for a variety
of equilibrium and dynamical models \cite{Andrade2005a}. This leads either
to exact results or to recurrence relations that can be numerically
iterated.

In this work, we analyze the avalanches of directed sandpile models on the
Apollonian network. We make use of properties of these specific network
and model to derive, in a first place, a series of exact results for the
distribution of avalanches. Then, these results can be extended, with the
help of the numerical iteration of the obtained recurrence relations, to
illustrate more general situations. More precisely, we are able to
investigate the fine details of the local mass flux, deriving the
appropriate multifractal spectra that describe the scaling properties of
the flux.

This work is organized as follows: In Section II we introduce our model,
discussing the role played by the number of levels $q$, in the Apollonian
hierarchy, that limit which nodes can receive mass from a toppling
neighbor. We also derive the basic expressions for the two and three-point
correlation functions that allow for the derivation of the local and total
fluxes. Results for the total flux, obtained by numerical iteration, are
discussed in Section III, for $1\leq q\leq 6$ . They are then compared
with analytical expressions derived for the $q=1$ and $q=2$. In Section
IV, a multifractal approach is used to present the scaling properties of
the flux for the distinct values of $q$. Finally, Section V closes the
paper with our concluding remarks.

\section{Directed sandpile models on the Apollonian network}

\begin{figure*}
\includegraphics[width=0.22\textwidth]{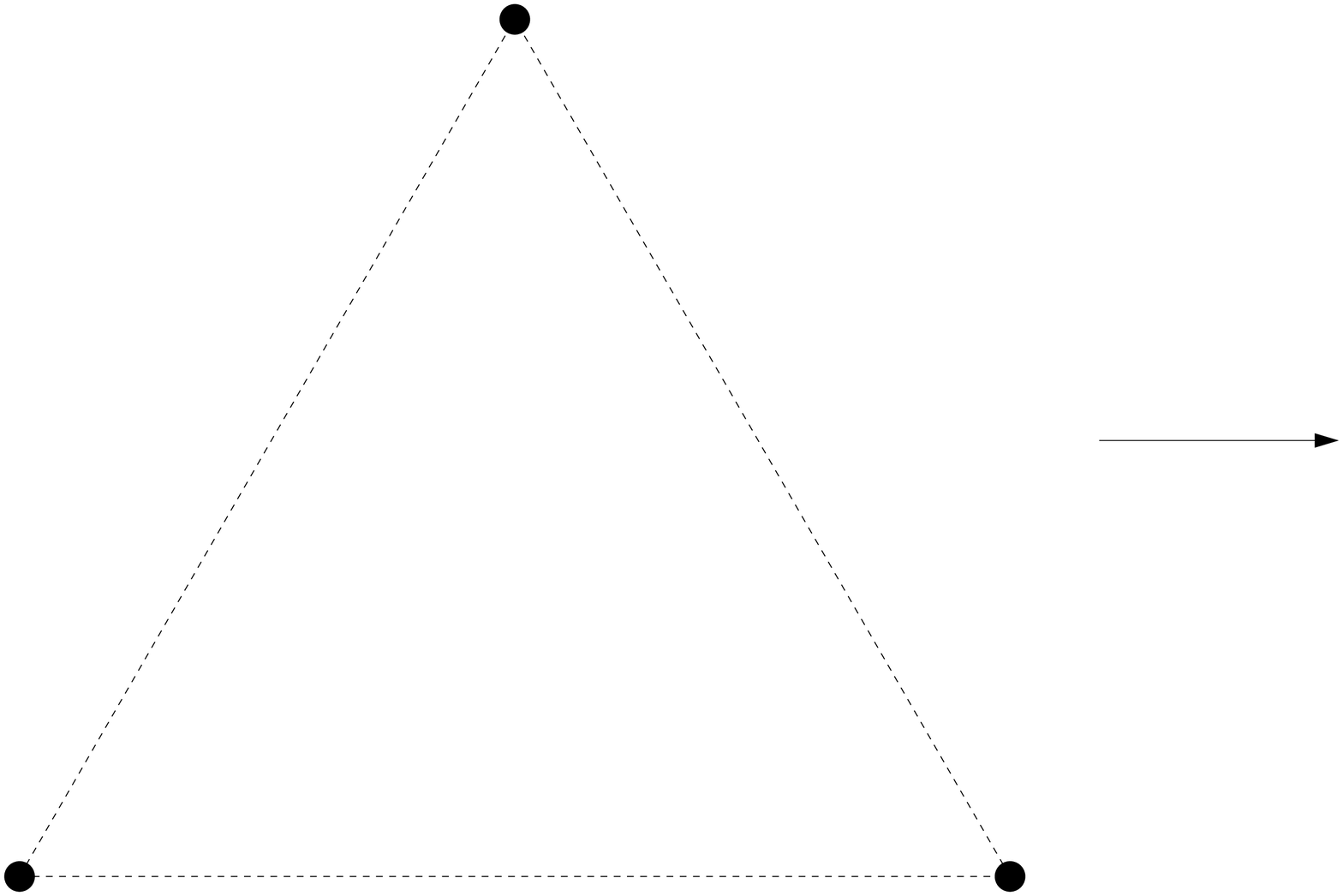}
\includegraphics[width=0.22\textwidth]{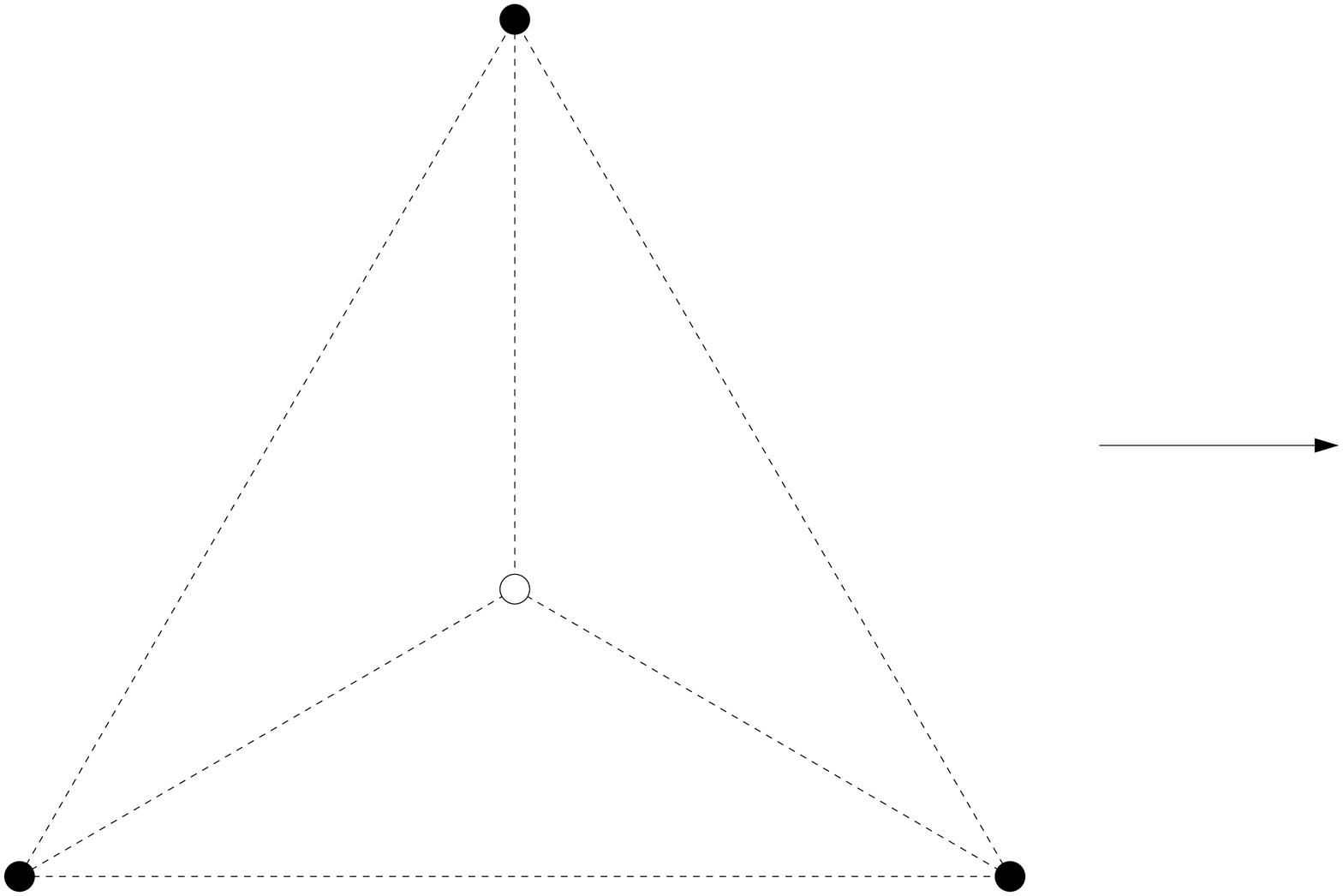}
\includegraphics[width=0.22\textwidth]{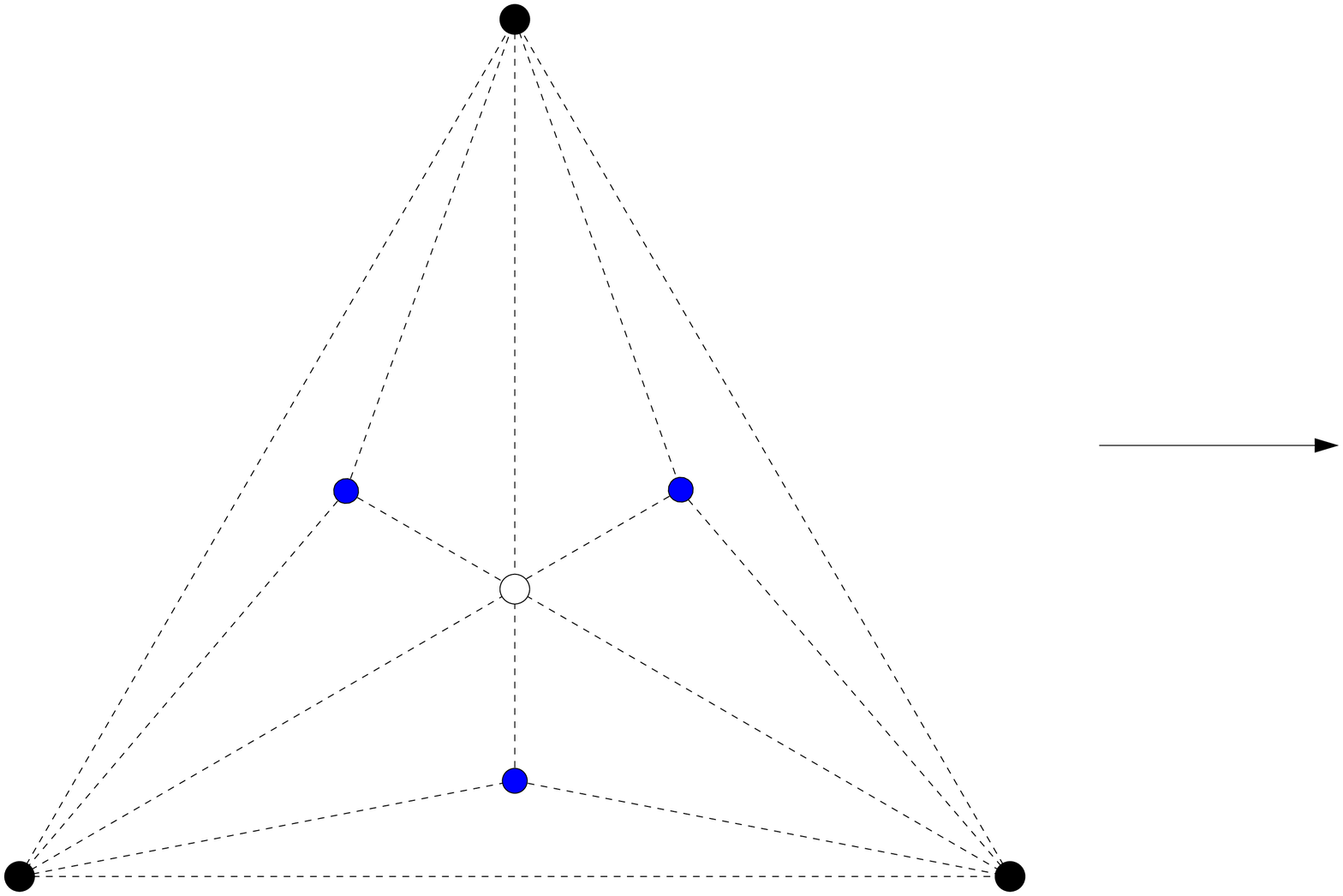}
\includegraphics[width=0.22\textwidth]{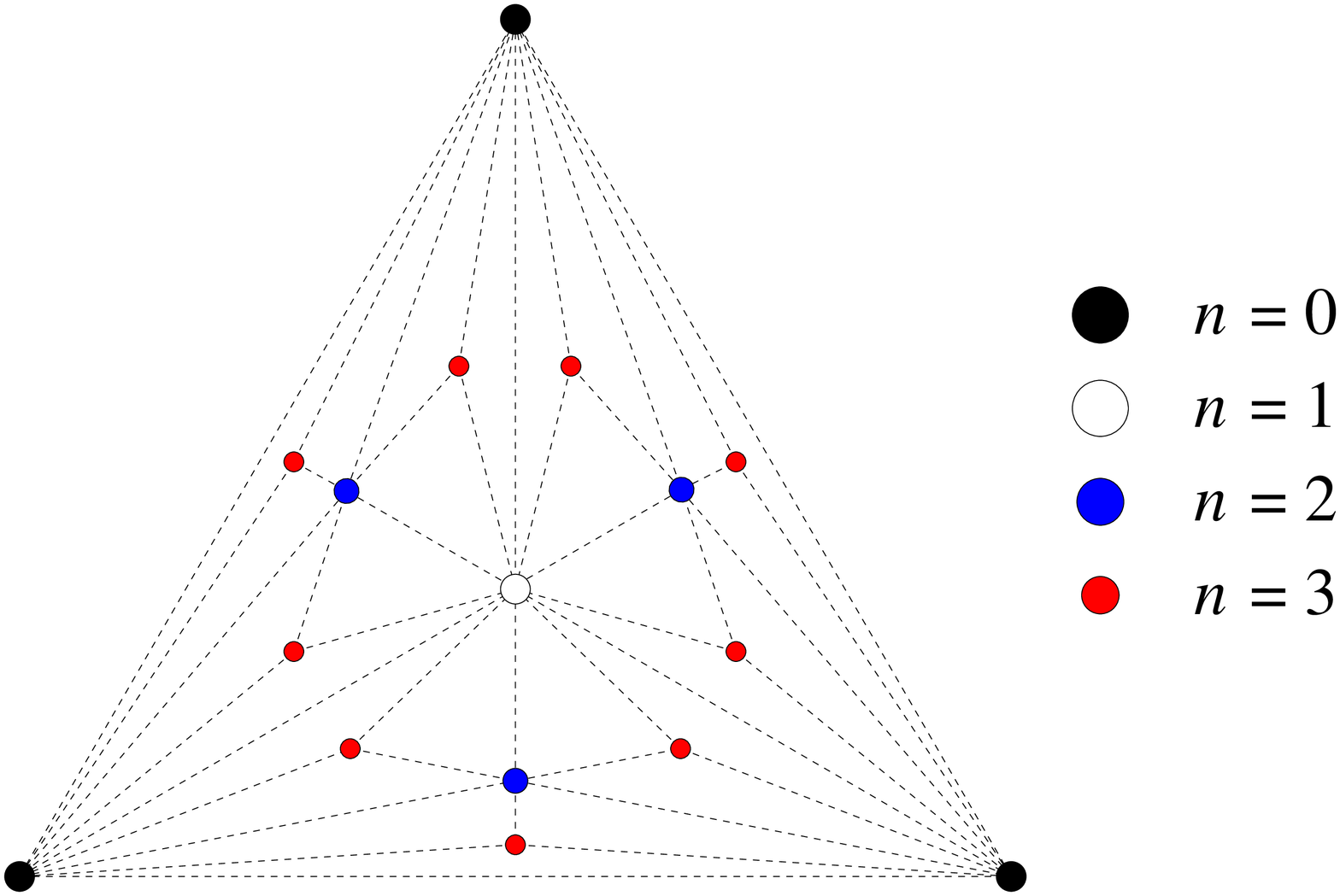}
\caption{\label{fig:Building-process-of}Building process of the Apollonian
network.}
\end{figure*}

The planar Apollonian network \cite{Andrade2005} is obtained from
the classical Apollonian space-filling packing of circles \cite{Boyd1973},
by associating nodes with the centers of the circles, and drawing
edges between nodes corresponding to pairs of touching circles. This
iterative building process is illustrated in Fig. \ref{fig:Building-process-of}.

\label{sec:The-directed-sandpile}The directed sandpile model of Dhar and
Ramaswamy \cite{Dhar1989} associates with each site $\mathbf{x}$ of a
hypercubic lattice a height variable $z\left(\mathbf{x}\right)$, which is
increased by $1$ when a grain is added to $\mathbf{x}$. If
$z\left(\mathbf{x}\right)$ exceeds a critical value $z_{c}$, the site
topples, and the height variables at the $\ell$ nearest neighbors of
$\mathbf{x}$ along a preferred direction increase by $1$, while
$z\left(\mathbf{x}\right)$ decreases by $\ell$. Without loss of
generality, $z_{c}$ can be chosen to equal $\ell$. The existence of a
preferred direction is essential to the exact solvability of the model,
not only in its original form but also in generalized versions
\cite{Andrade2003}.

\begin{figure}
\includegraphics[width=0.75\columnwidth]{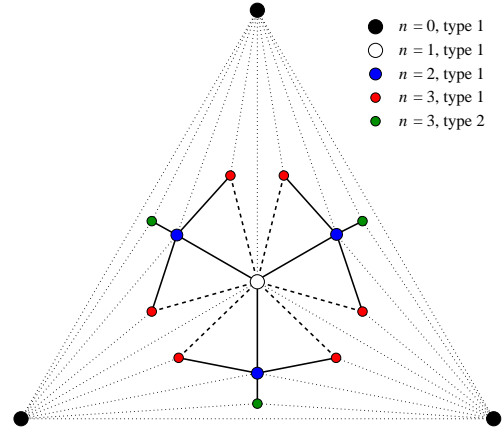}
\caption{\label{cap:Apollonian2}Apollonian network with $q=2$. Dotted
lines correspond to inactive connections; thick lines indicate connections
between sites in adjacent layers, while dashed lines connect sites
separated by $2$ layers. Note that there are two types of sites in layer
$n=3$. If $q=1$, also dashed lines become inactive.}
\end{figure}

In the Apollonian network, the building process offers an obvious choice
of a preferred direction. We define the $n$th layer of the network as the
set of sites added in the $n$th iteration of the process, and we postulate
that, when a site at a given layer topples, only sites in subsequent
layers can receive grains. However, the Apollonian network has the
peculiar property that each site in a given layer is connected to at least
one site in each subsequent layer. Thus, in the thermodynamic limit, any
site has an infinite number of neighbors in subsequent layers, leading to
an infinite critical height. In order to obtain a finite value of $z_{c}$,
we impose the restriction that only neighbors in the first $q$ subsequent
layers can receive grains when a site topples, the remaining connections
being inactive; see Fig.
\ref{cap:Apollonian2}. This leads to a $q$-dependent value of the critical
height $z_{c}$, which is the same for all sites in the network, provided
we forbid the addition of grains to the sites in the original triangle
(layer $n=0$). (For $q$ from $1$ to $6$, we have $z_{c}=3,$ $9$, $21$,
$45$, $93$, and $189$.) Thus, all allowed sites have an equivalent set of
neighbors in their subsequent layers, and we can study the properties of
avalanches by choosing any reference site $\mathbf{x}_{0}$. For
convenience, we choose $\mathbf{x}_{0}$ to be the site located at the
geometrical
center of the network (layer $n=1$). %

As in the original directed sandpile model, we define a two-point
correlation function $G_{0}\left(\mathbf{x};\mathbf{x}_{0}\right)$ which
measures the probability that a site $\mathbf{x}$ topples in the SOC state
due to an avalanche originated by adding a grain at $\mathbf{x}_{0}$.
Since the probability that a site topples, provided that $r$ of its
backwards neighbors have toppled, is equal to $r/z_{c}$, $G_{0}$ obeys the
recursion equation
\begin{equation}
G_{0}\left(\mathbf{x};\mathbf{x}_{0}\right)=\frac{1}{z_{c}}\left[\left.\sum_{\mathbf{y}}\right.^{\prime}G_{0}\left(\mathbf{y};\mathbf{x}_{0}\right)+\delta_{\mathbf{x},\mathbf{x}_{0}}\right],
\label{eq:recG0}
\end{equation}
with the primed summation running over all sites from which $\mathbf{x}$
can receive grains, according to the $q$-layer rule. Since
\begin{equation}
G_{0}\left(\mathbf{x}_{0};\mathbf{x}_{0}\right)=\frac{1}{z_{c}},
\end{equation}
the existence of a preferred direction allows us to solve Eq.
(\ref{eq:recG0}) for all $G_{0}\left(\mathbf{x};\mathbf{x}_{0}\right)$, at
least numerically.

The flux through the $n$th layer is given by
\begin{equation}
\phi\left(n\right)=\sum_{\mathbf{x}\in n}G_{0}\left(\mathbf{x};\mathbf{x}_{0}\right).
\label{eq:flux}
\end{equation}
Contrary to what is observed in hypercubic lattices, here
$\phi\left(n\right)$ generally depends on $n$, although it becomes
asymptotically constant for $n\gg1$, as we show below by numerical and
analytical calculations. If $m\left(n\right)$ is the average number of
sites in the $n$th layer that topple when at least one of them does, we
can write
\begin{equation}
\phi\left(n\right)=m\left(n\right)p\left(n\right),
\end{equation}
in which $p\left(n\right)$ is the probability that, in the SOC state, an
avalanche started at $\mathbf{x}_{0}$ reaches layer $n$.

If we assume that
\begin{equation}
p\left(n\right)\sim n^{-\alpha},
\end{equation}
with some exponent $\alpha$, the asymptotic constancy of
$\phi\left(n\right)$ allows us to conclude that
\begin{equation}
m\left(n\right)\sim\frac{1}{p\left(n\right)}\sim n^{\alpha}.
\end{equation}
Thus, the average mass of an avalanche reaching $n$ layers scales
as
\begin{equation}
M\left(n\right)=\sum_{t=1}^{n}m\left(t\right)\sim\int_{1}^{n}dt\
t^{\alpha}\sim n^{\alpha+1},
\end{equation}
and the probability that the total mass of an avalanche exceeds $M$
can be written as
\begin{equation}
\hat{p}\left(M\right)=p\left(n\left(M\right)\right)\sim
M^{-\frac{\alpha}{1+\alpha}}.
\end{equation}
Finally, we obtain for $\rho(M)$, the probability distribution of
avalanches with size $M$,
\begin{equation}
\rho(M)=\frac{d \hat{p}\left(M\right)}{dM}\sim
M^{-\frac{1+2\alpha}{1+\alpha}}\equiv M^{-\tau} .
\end{equation}

The exponent $\alpha$ can be calculated from the mean square flux
\begin{equation}
\Phi\left(n\right)=\left[m\left(n\right)\right]^{2}p\left(n\right)\sim
n^{\alpha},
\end{equation}
which is related to the three-point correlation function
$G\left(\mathbf{x}_{1},\mathbf{x}_{2};\mathbf{x}_{0}\right)$, defined as
the probability that sites $\mathbf{x}_{1}$ and $\mathbf{x}_{2}$, both in
the same layer, topple due to an avalanche started by adding a grain at
site $\mathbf{x}_{0}$. Explicitly, we have
\begin{equation}
\Phi\left(n\right)=\sum_{\mathbf{x}_{1},\mathbf{x}_{2}\in n}
G\left(\mathbf{x}_{1},\mathbf{x}_{2};\mathbf{x}_{0}\right).
\label{eq:flux2}
\end{equation}

As in the case of $G_{0}\left(\mathbf{x};\mathbf{x}_{0}\right)$,
we can write for $G\left(\mathbf{x}_{1},\mathbf{x}_{2};\mathbf{x}_{0}\right)$
a recursion equation,
\begin{equation}
G\left(\mathbf{x}_{1},\mathbf{x}_{2};\mathbf{x}_{0}\right)=\frac{1}{z_{c}^{2}}\left.\sum_{\mathbf{y}_{1},\mathbf{y}_{2}}\right.^{\prime}G\left(\mathbf{y}_{1},\mathbf{y}_{2};\mathbf{x}_{0}\right),
\end{equation}
with the primed summation running over all sites from which
$\mathbf{x}_{1}$ or $\mathbf{x}_{2}$ can receive grains, according to the
$q$-layer rule. This last equation can be solved by the \emph{Ansatz}
\cite{Dhar1990, Dhar1990a}
\begin{equation}
G\left(\mathbf{x}_{1},\mathbf{x}_{2};\mathbf{x}_{0}\right)=\sum_{\mathbf{y}}f\left(\mathbf{y};\mathbf{x}_{0}\right)G_{0}\left(\mathbf{x}_{1};\mathbf{y}\right)G_{0}\left(\mathbf{x}_{2};\mathbf{y}\right),
\label{eq:Ansatz}
\end{equation}
with the function $f\left(\mathbf{y};\mathbf{x}_{0}\right)$ determined
by the condition
\begin{equation}
G\left(\mathbf{x},\mathbf{x};\mathbf{x}_{0}\right)=G_{0}\left(\mathbf{x};\mathbf{x}_{0}\right),
\end{equation}
which leads to
\begin{equation}
\sum_{\mathbf{y}}f\left(\mathbf{y};\mathbf{x}_{0}\right)G_{0}\left(\mathbf{x};\mathbf{y}\right)G_{0}\left(\mathbf{x};\mathbf{y}\right)=G_{0}\left(\mathbf{x};\mathbf{x}_{0}\right).
\label{eq:condition}
\end{equation}
Summing over all sites $\mathbf{x}$ in the same layer $n$, using Eq.
(\ref{eq:flux}) and the fact that
$G\left(\mathbf{x};\mathbf{y}\right)=G\left(\mathbf{x}-\mathbf{y}+\mathbf{x}_{0};\mathbf{x}_{0}\right)$,
we can rewrite Eq. (\ref{eq:condition}) as
\begin{equation}
\sum_{t=1}^{n}F\left(t\right)K\left(n-t+1\right)=\phi\left(n\right),
\label{eq:FK}
\end{equation}
in which
\begin{equation}
F\left(t\right)=\sum_{\mathbf{y}\in
t}f\left(\mathbf{y}\right)\quad\mbox{and}\quad
K\left(t\right)=\sum_{\mathbf{x}\in
t}G_{0}\left(\mathbf{x};\mathbf{x}_{0}\right)G_{0}\left(\mathbf{x};\mathbf{x}_{0}\right).
\end{equation}

Starting from $n=1$, Eq. (\ref{eq:FK}) can be solved recursively
for $F\left(n\right)$. By substituting Eq. (\ref{eq:Ansatz}) into Eq. (\ref{eq:flux2}),
we can express $\Phi\left(n\right)$ in terms of $F\left(n\right)$,
\begin{equation}
\Phi\left(n\right)=\sum_{t=1}^{n}F\left(t\right)\left[\phi\left(n-t+1\right)\right]^{2}.
\end{equation}
The scaling behavior of $\Phi\left(n\right)$ determines the exponent
$\alpha$.

The case $q=1$ is immediately solved. In this limit, the Apollonian
network (with the three original vertices removed) reduces to a Cayley
tree with coordination number equal to $4$, as shown in Fig.
\ref{cap:Apollonian2}. The two-point correlation is easily seen to satisfy
\begin{equation}
G_{0}\left(\mathbf{x};\mathbf{x}_{0}\right)=\frac{1}{3^{n}},\qquad\mathbf{x}\in n,
\end{equation}
so that the average flux is $\phi\left(n\right)=1/3, \forall n,$ leading
to
\begin{equation}
K\left(n\right)=\frac{1}{3^{n+1}},\quad F\left(1\right)=3,\
F\left(n\right)=2\ \left(n>1\right).
\end{equation}
Thus, the mean square flux is given by
\begin{equation}
\Phi\left(n\right)=\frac{1}{9}+\frac{2}{9}n,
\end{equation}
corresponding to $\alpha=1 (\tau=3/2)$, characteristic of the mean-field
behavior associated with the directed model in Bravais lattices with
dimension $d\geq4$. For the purpose of comparison, the corresponding exact
values in $d=2$ are $\alpha=1/2, \tau=4/3.$


In the next two sections, we discuss the properties of the model for
$q>1$.

\section{Average behavior}

\label{sec:Average-behavior}For $q\ge2$, sites in the same layer are no
longer equivalent, since we are preserving the underlying topology of the
Apollonian network as defined by the building rule. Instead, those sites
are naturally grouped in different classes, defined by the structure of
their connection to sites in previous layers. In principle, this makes the
model amenable to analytical treatment. As we show in Appendix
\ref{sec:Analytical-treatment-for-q=3D2}, the analysis for $q=2$ is
already somewhat intricate, but it lends support to a series of
conclusions we obtain by numerical calculations. These are performed by
building an Apollonian network with up to $16$ layers (corresponding to
$21\ 523\ 363$ sites), imposing the $q$-layer rule, and solving
recursively Eqs. (\ref{eq:recG0}) and (\ref{eq:FK}). From this, we can
calculate both the mean flux $\phi\left(n\right)$ and the mean square flux
$\Phi\left(n\right)$ as functions of the layer index $n$. (In Sec.
\ref{sec:Multifractal-properties-of} we study the local properties of the
flux.)

\begin{figure}
\subfigure[]{\label{fig:Mean-flux}
\includegraphics[width=0.99\columnwidth]{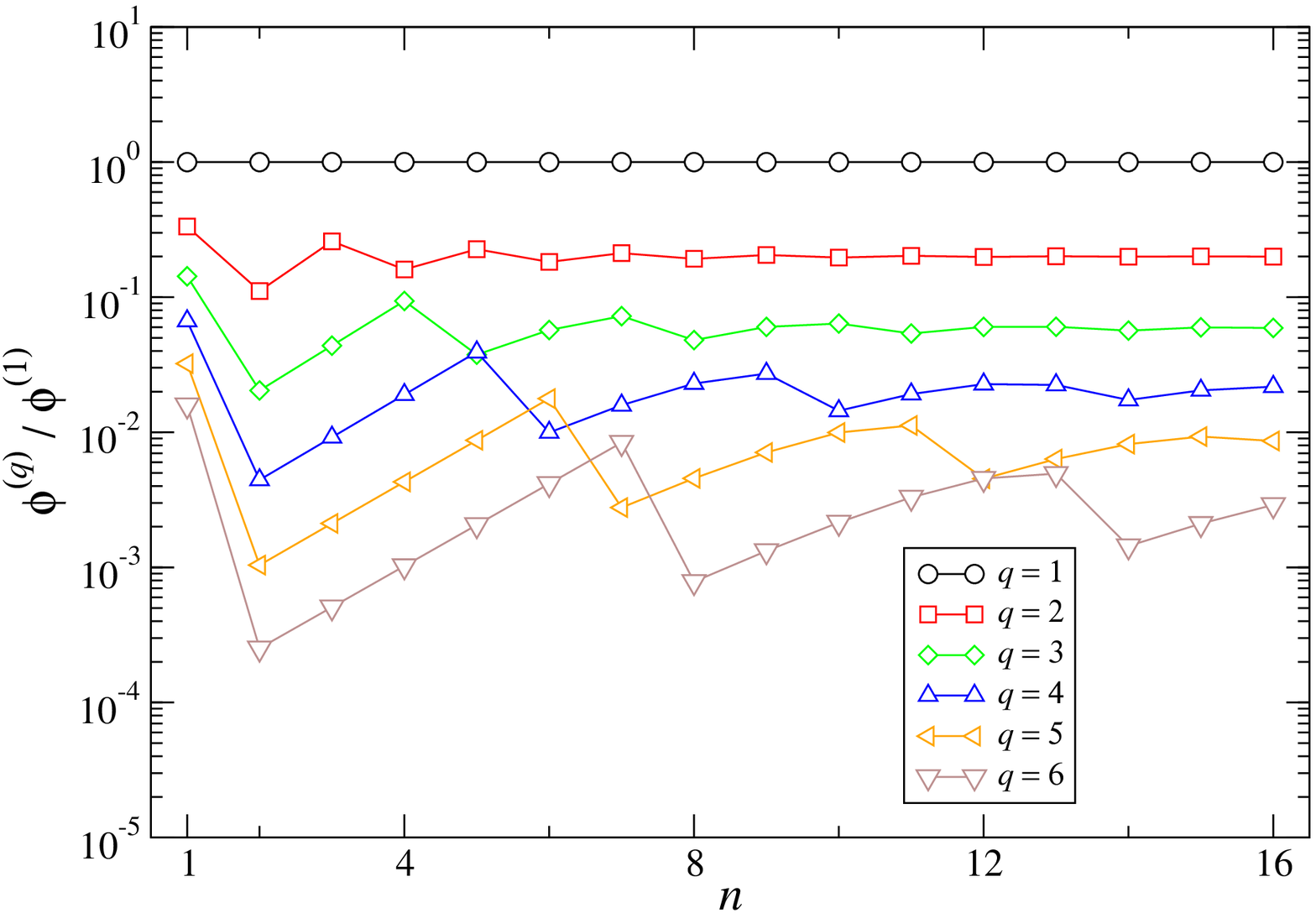}}
\subfigure[]{\label{fig:Mean-square-flux}
\includegraphics[width=0.99\columnwidth]{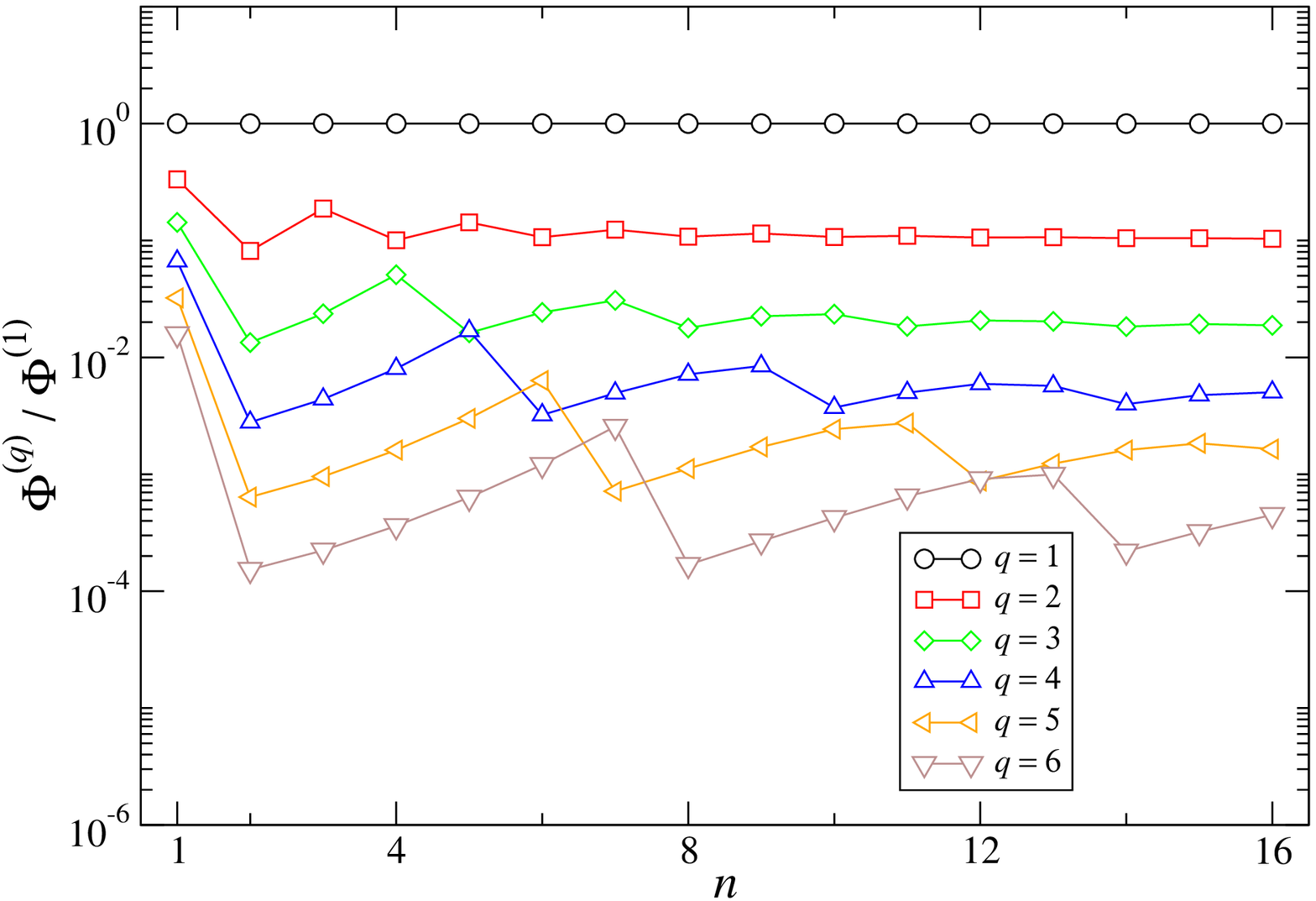}}
\caption{(a) Mean flux as a function of the layer index $n$, for different
values of $q$, divided by the mean flux for $q=1$. (b) Corresponding
curves for the mean square flux.}
\end{figure}

The first conclusion to emerge from our numerical analysis is that the
mean flux $\phi\left(n\right)$ becomes asymptotically constant for large
$n$, as already mentioned in Sec. \ref{sec:The-directed-sandpile}. This is
evident in Fig. \ref{fig:Mean-flux}, where we plot, for several values of
$q$, the ratio between $\phi\left(n\right)$ and the corresponding
(constant) result for $q=1$. Notice the oscillations in
$\phi\left(n\right)$ for small values of $n$. These are related to the
fact that, as the number of neighbors of a site in a given subsequent
layer increases with the layer index, so does the fraction of grains
received by each layer when the site topples. Since we have a single site
through which grains enter the system (in layer $n=1$), we note that, for
$q>1$, the mean flux reaching layer $n=2$ drops in comparison with the
total flux, then increases from $n=2$ to $n=q+1$, dropping again for
$n=q+2$. But this second drop is smaller, because sites in that layer
receive grains from all $q$ previous layers. As a result of this process,
the oscillations are smoothed out for sufficiently high values of
$n$. %

Corresponding curves for the mean square flux are shown in Fig.
\ref{fig:Mean-square-flux}. Again the curves oscillate for small values of
the layer index $n$, but approach a constant value for large $n$, showing
that $\Phi\left(n\right)$ always satisfies the scaling
form\begin{equation} \Phi\left(n\right)\sim n.\end{equation} This means
that asymptotically the distribution of avalanche sizes follows a
power-law with exponent $\alpha=1$, irrespective of $q$.

This prevalence of a mean-field behavior could be anticipated on the basis
of the tree-like topology of the lattice obtained by imposing the
$q$-layer rule. A similar situation arises in other sandpile models on
different forms of decorated Cayley trees \cite{Dhar1990a,Papoyan1995};
since the correlation length is infinite in the SOC state, the mean-field
behavior characteristic of the ordinary Cayley tree (or more precisely the
Bethe lattice) is recovered.

In the Apollonian network, the $q$-layer rule defines a typical length
beyond which average properties become indistinguishable from those
of the model on a Cayley tree. However, at smaller scales, hints of
the behavior corresponding to the genuine Apollonian network do appear,
for instance in the oscillations observed in the mean-flux curves.
As clearly shown in Fig. \ref{fig:Mean-flux}, $\phi\left(n\right)$
depends exponentially on $n$ between $n=2$ and $n=q+1$,\begin{equation}
\phi\left(n\right)=Ae^{an},\end{equation}
with a $q$-dependent prefactor $A$, but a nearly constant value
of $a\simeq0.7$. The prefactor $A$ decreases exponentially with
$q$, since it is related to the inverse threshold height $1/z_{c}$.
The exponential (rather than linear) dependence of $\phi\left(n\right)$
is a consequence of the exponential increase in the number of neighbors
as a function of the layer separation. In the $q\rightarrow\infty$
limit, the central site topples only after the addition of an enormous
number of grains, most of which are then received by sites in very
distant layers. As a consequence, all avalanches have arbitrarily
large range.

\section{Multifractal properties of the flux}

\label{sec:Multifractal-properties-of}
\begin{figure}
\subfigure[]{\label{fig:Flux-distributions}
\includegraphics[width=0.99\columnwidth]{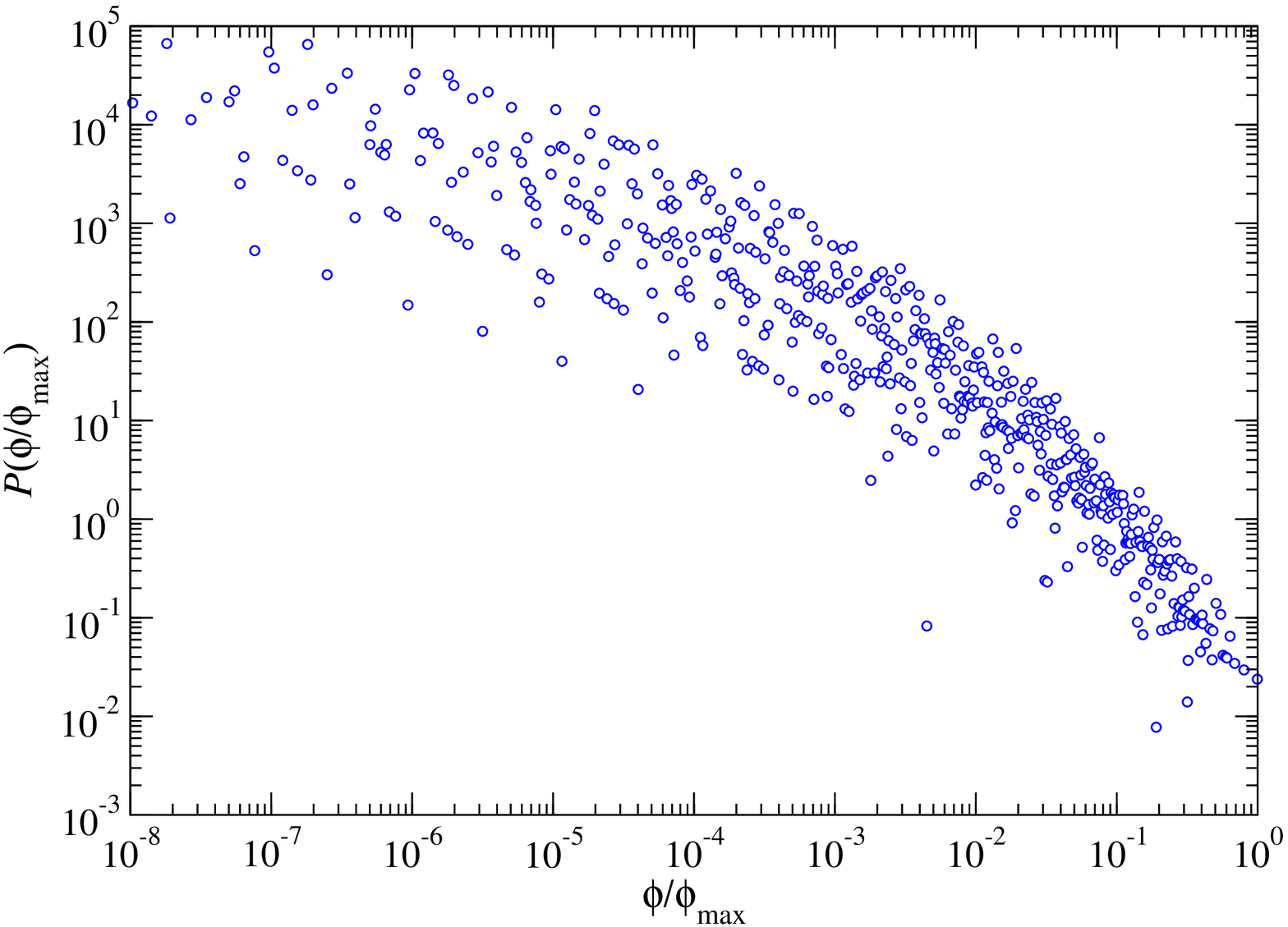}}
\subfigure[]{\label{fig:q3momentos}
\includegraphics[width=0.99\columnwidth]{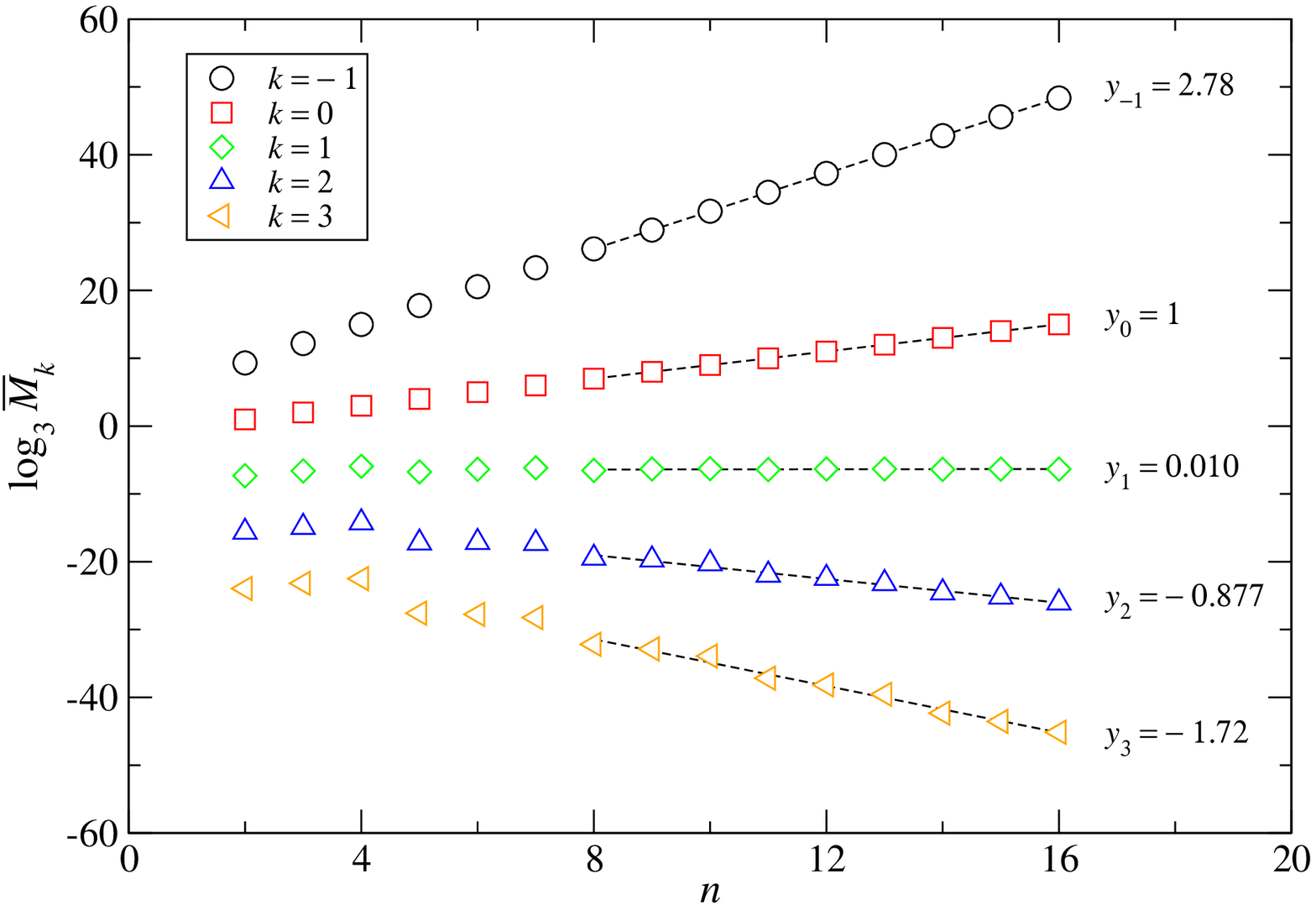}}
\caption{(a) Flux distributions in a given layer, re-scaled by the
corresponding maximum flux, for $q=2$ and $n=22$. (b) Moments of the flux
as a function of $n$, for $q=3$ and several values of $k$, with the
corresponding exponents $y_{k}$. Notice that there is no linear relation
between the exponents.}
\end{figure}

Although the average behavior of the
flux reproduces that of the mean-field limit, the local-flux distribution
reveals interesting properties already for $q=2$. In Fig.
\ref{fig:Flux-distributions} we plot histograms of the local flux $\phi$
for $n=22$. Note that, due to a precise identification of the distinct
types of sites for the $q=2$ model, we were able to consider much larger
number of nodes $(>10^{10})$ than for the results reported in the previous
section. The fluxes are re-scaled by the corresponding maximum flux
$\phi_{\mathrm{max}}$ in that layer. %

In analogy with studies on the distribution of currents in the incipient
infinite cluster of a random-resistor network \cite{Stauffer2003}, it is
interesting to evaluate the moments of the flux, in order to better reveal
the scaling properties hidden in Fig. \ref{fig:Flux-distributions}. So we
use the definition
\begin{equation}
\overline{M}_{k}\left(n\right)=\sum_{\mathbf{x}\in
n}\left(\frac{\phi_{\mathbf{x}}}{\phi_{0}}\right)^{k},
\end{equation}
in which the summation runs over all sites $\mathbf{x}$ in the $n$th
layer of the Apollonian network,
\begin{equation}
\phi_{\mathbf{x}}\equiv G_{0}\left(\mathbf{x};\mathbf{x}_{0}\right)
\end{equation}
and $\phi_{0}$ is the initial flux. It turns out that, for all real
values of $k$, the moments satisfy scaling relations given by
\begin{equation}
\overline{M}_{k}\left(n\right)\sim e^{u_{k}n},
\end{equation}
with well defined coefficients $u_{k}$, so that, in terms of the
system size
\begin{equation}
L\sim3^{n+1},
\end{equation}
we have
\begin{equation}
\overline{M}_{k}\left(L\right)\sim L^{y_{k}},
\end{equation}
with $y_{k}=u_{k}/\ln3$. For $q=1$, all sites in a given layer $n$ have
the same flux $3^{-n}$, so that the exponents $y_{k}$ are given by
$y_{k}=1-k$. For $q\geq2$, on the other hand, we see from our numerical
calculations that there is no simple linear relation between the exponents
$y_{k}$, suggesting that no single number characterizes the current
distribution. This is a signature of multifractal behavior. A plot of
$\overline{M}_{k}\left(n\right)$ for $q=3$ and several values of $k$ is
shown in Fig. \ref{fig:q3momentos}.

\begin{figure}
\includegraphics[width=0.99\columnwidth]{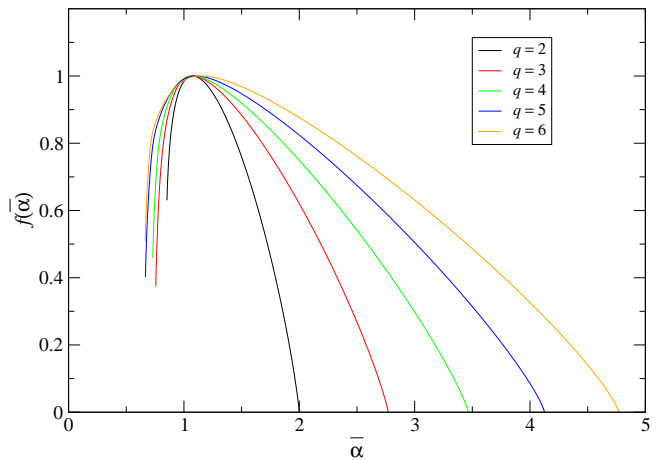}
\caption{\label{fig:fdealfa}Plots of $f\left(\overline{\alpha}\right)$ for
different values of $q$.}
\end{figure}

To further investigate the multifractal properties of the flux
distribution, we evaluate the multifractal spectrum
$f\left(\overline{\alpha}\right)$, defined by a Legendre transform of the
exponents $y_{k}$,
\begin{equation}
f\left(\overline{\alpha}\right)=y_{k}+k\overline{\alpha},\qquad\overline{\alpha}=-\frac{dy_{k}}{dk}.
\label{eq:falfa}
\end{equation}
Plots of $f\left(\overline{\alpha}\right)$ for several values of $q$ are
shown in Fig. \ref{fig:fdealfa}. Note that,  according to Eq.
(\ref{eq:falfa}), the maximum of $f\left(\overline{\alpha}\right)$ occurs
for the value of $\alpha$ associated with $k=0$, for which
$f\left(\overline{\alpha}_{k=0}\right)=y_{0}$. Indeed, for $q=1$ (not
shown in Fig. \ref{fig:fdealfa}), the curve consists of a single point,
$\left(\overline{\alpha}_{k=0},y_{0}\right)=\left(1,1\right)$,
corresponding to a monofractal behavior. Within numerical errors, that
point is the maximum of all curves, in agreement with the fact
that $y_{0}=1$ for all values of $q$.%

For $q\geq 2$, the left (right) end of the curves reflects the scaling
behavior of the set of points associated with the largest (smallest)
fluxes. Although not visible in the plots, the density of points is much
larger near the ends of the curves, with intermediate points coming mostly
from values of $k$ between $-1$ and $0$. The width of the curves increases
with $q$, presumably diverging as $q\rightarrow\infty$. This is related to
the fact that larger values of $q$ lead to a larger range of values of the
flux in each layer of the lattice.

\section{Conclusions}

In this work we investigate directed sandpile models on the Apollonian
network. Exact results were obtained for the avalanche distributions when
$q=1$ and 2. These correspond to the situations where an unstable node
topples only to neighbors introduced in the first and the second
subsequent generations, respectively. The avalanche distributions follow
power-law behavior, with typical mean field exponents. The iteration of
the exact expression for the two and three point correlation function
provides evidences for the same asymptotic behavior, regardless of the
finite value of $q$. On the other hand, our results also show the
emergence of large oscillatory deviations due to finite size effects. This
general behavior can be explained by noting that any finite value of $q$
asymptotically constraints the sandpile model on the Apollonian network to
the structure of a tree, where it behaves like a mean field model.

The investigation of the local properties of the fluxes through each node
shows that the network geometry induces a large degree of inhomogeneity in
the sandpile model. This effect has not been observed for the same model
in Euclidian lattices. Nevertheless, this dependence can be accurately
accounted for by a multifractal analysis. Only for $q=1$, when the model
is equivalent to that defined on a Cayley tree, all nodes become
indistinguishable, and the scaling analysis reduces to a single point.

Finally, the comparison with results found for another sandpile model on
scale-free networks \cite{Goh2003} shows similarities, in the mean field
behavior when all nodes share the same critical height or the critical
height depends locally on the node degree.

\appendix

\section{Analytical treatment for $q=2$}

\label{sec:Analytical-treatment-for-q=3D2}

For $q\geq2$, the sites in each layer of the Apollonian network can
be grouped in types, according to how they are connected to their
backwards neighbors. This feature can be exploited in order to obtain
analytical results for the behavior of the directed sandpile model.
Here we deal with the case $q=2$, which allows us to check our numerical
results in a reasonably simple way. It is clear that the treatment
can be extended to higher values of $q$, with basically the same
results, but a considerably larger amount of work.%

Layer $n=1$ of the network contains only one site, while the three
sites in layer $n=2$ are all equivalent. However, already for $n=3$
two types of sites are present: sites of type $1$ receive grains
from sites in the two previous layers, while sites of type 2 receive
grains only from the latter layer; see Fig. \ref{cap:Apollonian2}.
For $n=4$, two additional types of sites would appear, since it is
possible that a site receives grains from sites of types $1$ or $2$,
in one or two of the previous layers. It is easy to convince oneself
that the number of site types doubles for each additional layer (starting
at $n=2$), and that the types can be labeled so that each site of
type $s$ has as nearest neighbors in the next layer two sites of
type $2s-1$ and one site of type $2s$.

Denoting by $g_{n,s}$ the value of $G_{0}\left(\mathbf{x};\mathbf{x}_{0}\right)$
for a site $\mathbf{x}$ of type $s$ in layer $n$, and by $\nu_{n,s}$
the number of such sites, the flux through layer $n$ can be written
as
\begin{equation}
\phi\left(n\right)=\sum_{s=1}^{2^{n-2}}\nu_{n,s}g_{n,s}.\label{eq:fng}
\end{equation}
In order to estimate $\phi\left(n\right)$, we must investigate the
asymptotic behavior of both $\nu_{n,s}$ and $g_{n,s}$.

Our choice of labels allows us to write, for $s=2j-1$ ($j=1$, $2$,
$3$, $\ldots$ ),
\begin{equation}
g_{n,2j-1}=\frac{1}{9}\left(g_{n-1,j}+g_{n-2,\left\lfloor
\frac{j+1}{2}\right\rfloor
}\right),\quad\nu_{n,2j-1}=2\nu_{n-1,j},\label{eq:g1}
\end{equation}
in which $\left\lfloor w\right\rfloor $ denotes the integer part
of the number $w$, and, for $s=2j$,
\begin{equation}
g_{n,2j}=\frac{1}{9}g_{n-1,j},\quad\nu_{n,2j}=\nu_{n-1,j}.\label{eq:g2}
\end{equation}

Equations (\ref{eq:g1}) and (\ref{eq:g2}), being recursive expressions,
can be solved numerically to yield all $g_{n,s}$ and $\nu_{n,s}$
in terms of $g_{1,1}$ and $\nu_{1,1}$. However, analytical results
can be derived from the observation that $g_{n,1}$ behaves as
\begin{equation}
g_{n,1}\sim\zeta^{n},
\end{equation}
with $\zeta=\left(1+\sqrt{37}\right)/18\simeq0.393$ being determined from
the solution of the equation
\begin{equation}
\zeta^{2}=\frac{1}{9}\left(\zeta+1\right).
\end{equation}
Consequently, $g_{n,s}$ satisfies
\begin{equation}
g_{n,s}\simeq A_{s}\zeta^{n}, \label{eq:gns}
\end{equation}
with constant prefactors $A_{s}$. Moreover, the multiplicities $\nu_{n,s}$
are such that $\nu_{n,1}=3\cdot2^{n-2}$ ($n\geq2$) and the ratios
$f_{s}\equiv\nu_{n,s}/\nu_{n,1}$ satisfy
\begin{equation}
\left\{ \begin{array}{c}
f_{2j-1}=\frac{\nu_{n,2j-1}}{\nu_{n,1}}=\frac{2\nu_{n,j}}{2\nu_{n,1}}=f_{j},\\
f_{2j}=\frac{\nu_{n,2j}}{\nu_{n,1}}=\frac{\nu_{n,j}}{2\nu_{n,1}}=\frac{1}{2}f_{j}.
\end{array}\right.
\label{eq:ff}
\end{equation}

We can rewrite Eq. (\ref{eq:fng}) as
\begin{equation}
\phi\left(n\right)=\nu_{n,1}\sum_{s=1}^{2^{n-2}}f_{s}g_{n,s}=
\nu_{n,1}\sum_{m=0}^{n-2}\Gamma_{n,m},
\end{equation}
with
\begin{equation}
\Gamma_{n,0}=g_{n,1}\quad\textrm{and}\quad\Gamma_{n,m}
=\sum_{s=1+2^{m-1}}^{2^{m}}f_{s}g_{n,s}\quad\left(m\geq1\right).
\end{equation}
Making use of the definition of $\Gamma_{n,m}$ and of Eqs. (\ref{eq:g1}),
(\ref{eq:g2}) and (\ref{eq:ff}), we can obtain the recursion
equation
\begin{equation}
\Gamma_{n,m}=\frac{1}{6}\left(\Gamma_{n-2,m-2}+\Gamma_{n-1,m-1}\right).
\label{eq:Gammanm}
\end{equation}
Keeping in mind Eq. (\ref{eq:gns}), we expect that $\Gamma_{n,m}$ takes
the asymptotic form
\begin{equation}
\Gamma_{n,m}\simeq\gamma_{m}\zeta^{n}.
\end{equation}

Substituting this last expression into Eq. (\ref{eq:Gammanm}), we
conclude that the constants $\gamma_{m}$ satisfy the equation
\begin{equation}
\gamma_{m}\zeta^{2}-\frac{1}
{6}\gamma_{m-1}\zeta-\frac{1}{6}\gamma_{m-2}=0,
\end{equation}
which can be solved by $\gamma_{m}\simeq\gamma_{0}\theta^{m}$, with
$\theta=\left(2\zeta\right)^{-1}$. We then have
\begin{equation}
\Gamma_{n,m}\sim\theta^{m}\zeta^{n}=\frac{1}{2^{m}}\zeta^{n-m},
\end{equation}
and the flux $\phi\left(n\right)$ scales with the layer index as
\begin{equation}
\phi\left(n\right)\sim2^{n}\cdot\zeta^{n}\sum_{m=0}^{n-2}\theta^{m}
=\left(2\zeta\right)^{n}\frac{\theta^{n-1}-1}{\theta-1}.
\end{equation}
Since $\theta>1$, this is equivalent to
\begin{equation}
\phi\left(n\right)\sim\left(2\zeta\theta\right)^{n}=1,
\end{equation}
so that the flux becomes asymptotically constant for $n\gg1$.

The function $K\left(n\right)$, defined by
\begin{equation}
K\left(n\right)=\sum_{\mathbf{x}\in n}G_{0}\left(\mathbf{x};
\mathbf{x}_{0}\right)G_{0}\left(\mathbf{x};
\mathbf{x}_{0}\right)=\sum_{s=1}^{2^{n-2}}\nu_{n,s}g_{n,s}^{2},
\end{equation}
scales as $K\left(n\right)\sim\zeta^{n}$, and thus vanishes exponentially
for large $n$. In the same limit, the function $F\left(n\right)$, related
to $K\left(n\right)$ and $\phi\left(n\right)$ through the equation
\begin{equation}
\sum_{t=1}^{n}F\left(t\right)K\left(n-t+1\right)=\phi\left(n\right),
\end{equation}
tends to a constant value. As a consequence, the mean square flux must
scale as
\begin{equation}
\Phi\left(n\right)=
\sum_{t=1}^{n}F\left(n\right)\left[\phi\left(n-t+1\right)\right]^{2}\sim
n,
\end{equation} yielding the mean-field
exponent $\alpha=1$.

\acknowledgments

This work was partially financed by the Brazilian agencies CNPq, FUNCAP,
and FAPESB. HJH acknowledges the support of the Max Planck prize.


\end{document}